\documentclass[sigconf]{acmart}
\usepackage{graphicx} 
\usepackage{algorithm}
\usepackage{algpseudocode}
\usepackage{subcaption}
\usepackage{wrapfig}

\newcommand{\cn}[0]{\textbf{\color{blue} CITE}}

\newcommand{\hg}[1]{\textcolor{purple}{HG: #1}}

\newcommand{\db}[0]{Turso}
\newcommand{\tool}[0]{\textsc{DIRT}}

\usepackage[utf8]{inputenc}
\usepackage[english]{babel}

\usepackage{xcolor}
\definecolor{dkblue}{rgb}{0,0.1,0.7}
\definecolor{ltblue}{rgb}{0,0.1,0.7}
\definecolor{dkbluet}{rgb}{0,0.1,0.7}
\definecolor{ltbluet}{rgb}{0.8,0.8,1}
\definecolor{dkgreen}{rgb}{0,0.5,0}
\definecolor{ltgreen}{rgb}{0.8,1,0.8}
\definecolor{dkred}{rgb}{0.7,0,0}
\definecolor{ltred}{rgb}{1,0.8,0.8}
\definecolor{dkviolet}{rgb}{0.3,0,0.5}
\definecolor{ltviolet}{rgb}{1,0.8,1}
\definecolor{dkpurple}{rgb}{0.7,0,0.4}
\definecolor{ltpurple}{rgb}{1,0.8,1}
\definecolor{dkorange}{rgb}{0.8,0.4,0}
\definecolor{ltorange}{rgb}{1,0.8,0.6}
\definecolor{dkyellow}{rgb}{0.6,0.6,0}
\definecolor{ltyellow}{rgb}{1,1,0.6}
\definecolor{dkgray}{rgb}{0.4,0.4,0.4}
\definecolor{ltgray}{rgb}{0.9,0.9,0.9}
\definecolor{olive}{rgb}{0.4, 0.4, 0.0}
\definecolor{dkolive}{rgb}{0.4, 0.4, 0.0}
\definecolor{ltolive}{rgb}{0.9, 0.9, 0.0}
\definecolor{teal}{rgb}{0.0,0.5,0.5}
\definecolor{azure}{rgb}{0.0, 0.4, .8}
\definecolor{ltazure}{rgb}{0.8, 0.9, 1}
\definecolor{dkazure}{rgb}{0.0, 0.4, .8}

\usepackage{listings}

\lstdefinelanguage{GA}{ 
    mathescape=true,
    texcl=false, 
    morekeywords=[1]{gen, genOf, pick, assert, assume},
    morekeywords=[2]{property, expression, row, id, columntype},
    morekeywords=[3]{SELECT, INSERT, FROM, WHERE, UNION, ALL, DELETE, CREATE, TABLE, INDEX, INTO, VALUES, UPDATE, SET},
    morekeywords=[4]{AND, NOT, NULL, GLOB, BETWEEN},
    morekeywords=[5]{by, done, exact, reflexivity, tauto, romega, omega,
        assumption, solve, contradiction, discriminate},
    morekeywords=[6]{do, last, first, try, idtac, repeat},
    morecomment=[l]{--},
    showstringspaces=false,
    morestring=[b]",
    morestring=[d]’,
    tabsize=3,
    extendedchars=false,
    sensitive=true,
    breaklines=false,
    basicstyle=\small,
    captionpos=b,
    columns=[l]flexible,
    identifierstyle={\ttfamily\color{black}},
    keywordstyle=[1]{\ttfamily\color{dkviolet}},
    keywordstyle=[2]{\ttfamily\color{dkgreen}},
    keywordstyle=[3]{\ttfamily\color{ltblue}},
    keywordstyle=[4]{\ttfamily\color{dkred}},
    keywordstyle=[5]{\ttfamily\color{dkred}},
    stringstyle=\ttfamily,
    commentstyle={\ttfamily\color{lightgray}},
    literate=
    {\\emp}{$\emptyset$}1
    {\\.}{$\cdot$}1
    {\\forall}{{\color{dkgreen}{$\forall\;$}}}1
    {\\exists}{{$\exists\;$}}1
    {<-}{{$\leftarrow\;$}}1
    {->}{{$\rightarrow\;$}}1
    {<->}{{$\leftrightarrow\;$}}1
    {<==}{{$\leq\;$}}1
    {\#}{{$^\star$}}1 
    {\\o}{{$\circ\;$}}1 
    {\@}{{$\cdot$}}1 
    {\/\\}{{$\wedge\;$}}1
    {\\\/}{{$\vee\;$}}1
    {~}{{$\sim$}}1
    {\@\@}{{$@$}}1
    {\\mapsto}{{$\mapsto\;$}}1
    {\\hline}{{\rule{\linewidth}{0.5pt}}}1
}[keywords,comments,strings]
\lstnewenvironment{gacode}{\lstset{language=GA,mathescape=true,basicstyle=\ttfamily\small}}{}

\newcommand{\ga}{\lstinline[language=GA, basicstyle=\ttfamily\small]}
\newcommand{\sql}{\lstinline[language=GA, basicstyle=\ttfamily\small]}

\setcopyright{none}
\renewcommand\footnotetextcopyrightpermission[1]{}

\title{\tool{}: Database-Integrated Random Testing}

\begin{document}

\author{Alperen Keles}
\affiliation{%
  \institution{University of Maryland, College Park}
  \city{College Park}
  \country{USA}
}
\email{akeles@umd.edu}

\author{Ethan Chou}
\affiliation{%
  \institution{University of Maryland, College Park}
  \city{College Park}
  \country{USA}
}
\email{echou1@umd.edu}

\author{Harrison Goldstein}
\affiliation{%
  \institution{University of Buffalo}
  \city{Buffalo}
  \country{USA}
}
\email{hgoldste@buffalo.edu}

\author{Leonidas Lampropoulos}
\affiliation{%
  \institution{University of Maryland, College Park}
  \city{College Park}
  \country{USA}
}
\email{leonidas@umd.edu}

\begin{abstract}

Database management systems (DBMSs) are notoriously complex, making them difficult to test effectively, especially during early development when many features are incomplete. 
Traditional testing tools like SQLancer and SQLSmith are highly effective for mature databases, but they struggle with high false positive rates and low actionability when applied to evolving systems.
We present \tool{}, a paradigm designed specifically for testing databases during development, which integrates a testing framework directly into the DBMS, enabling the random testing process to evolve in tandem with the system and reducing false positives by construction.
We introduce generation actions, an abstraction for allowing database developers rather than testing experts to specify correctness properties.
We evaluate \tool{} on \db{}, an actively developed SQLite-compatible OLTP engine, and show that it finds 23 unique, confirmed bugs--significantly outperforming off-the-shelf SQLancer variants in terms of true positive rate and usefulness of bug reports. Our results demonstrate that embedding testing infrastructure within the DBMS can dramatically improve its effectiveness and usability during development.

\end{abstract}
\maketitle  

\section{Introduction} \label{sec:intro}

Database Management Systems (DBMSs) are complex systems, and  inevitably, such complexity leads to bugs. To discover bugs, DBMSs have historically used random testing, with approaches ranging from binary fuzzing with AFL~\cite{AFLplusplus} to structured query generation in SQLSmith~\cite{sqlsmith}, and to the powerful test oracles of SQLancer~\cite{sqlancer}.
%
Within prior work, SQLancer is distinguished by using specialized test oracles that reveal logic bugs in DBMSs, and it has proven extremely successful at finding bugs in mature databases.

Unfortunately, off-the-shelf testing frameworks are not as effective at finding actionable bugs in DBMSs that are actively being developed and have many missing features. 
When the space of randomly generated inputs is larger than the system under test supports, many of the generated tests lie in the unimplemented portion of the system,
registering as false positives.
Naturally, such false counterexamples are of no help to developers of DBMSs.

One way to alleviate this issue is to tailor the testing framework to the DBMS under test, ensuring that the testing framework only tests implemented features. 
Unfortunately, existing frameworks are not built with this flexibility in mind. In an off-the-shelf tester, such tailoring would require developers of the DBMS to dive into the internals of an external project, modify it to suit their needs, and maintain those changes as the DBMS evolves.

We propose \emph{database-integrated testing}, an alternative testing paradigm for testing databases during development.
We embed the testing tool within the database itself, allowing for the generators
to naturally evolve alongside the DBMS, therefore avoiding false positives by construction. 
At the same time, we leverage this tight integration to empower developers to
write their own oracles by developing a domain-specific language (DSL) for expressing
properties customized to particular features being developed.
%



We study this problem in the context of \db{}, an open-source SQLite3-compatible OLTP database engine in active development. With an average of more than 30 commits a day, \db{} is a rapidly evolving target for testing. 
\tool{} helped find 23 confirmed bugs, all of which were subsequently fixed.
In comparison, running SQLancer-SQLite3 with minimal modifications in \db{} led to a false positive rate of 96.5\% with only one new bug, and running SQLancer-\db{}, an existing independently developed SQLancer integration for \db{}, without modifications led to a false positive rate of 58\% with six newly discovered bugs.

In summary, our contributions are as follows:
\begin{enumerate}
    \item We propose \tool{}, a paradigm for testing databases during development which
     integrates a testing framework within the database and provides convenient and flexible abstractions to developers.  
    \item We implement \tool{} for \db{}, an open source SQLite3-compatible OLTP database under active development.
    \item We evaluate the effectiveness of \tool{} by finding actionable bugs in \db{}: we found 23 unique confirmed bugs with very few false positives compared to the state-of-the-art.
\end{enumerate}

\section{Background and Related Work} \label{sec:background}

We begin by providing background: first on coverage-guided fuzzing and property-based testing at a high level, and then focusing on SQLancer, the prominent framework for testing databases.

\subsection{Fuzzing and Property-Based Testing} 
\label{subsec:fuzz-pbt}

%
%
In Fuzz Testing~\cite{fuzzingbook2024}, the primary focus is on smart and effective test-case generation that can work out of the box with minimal user inputs.
The key idea is to leverage {\em runtime feedback}, usually in the form of branch coverage~\cite{AFLplusplus,AFLFast}, though not always~\cite{FuzzFactory}, 
to keep track of inputs that exhibit interesting behavior (e.g. uncovered new paths) and then mutate them in the hopes of discovering even more interesting inputs.
The most common oracle for deciding if an input is a bug or not is simple: does the program crash? Other oracles include fuzzing against a model or a different implementation (known as differential fuzzing~\cite{pbtpractice, apollo}) or checking for excessive execution times~\cite{perffuzz}.

Property-Based Testing (PBT), on the other hand,  primarily focuses on empowering users to write their own oracles~\cite{quickcheck, htsi}. PBT frameworks usually offer DSLs for expressing oracles in the form of universally quantified executable predicates, as well as infrastructure
for writing {\em generators} - programs that generate test inputs to test such predicates~\cite{quickcheck, generatinggenerators}.
%
Although traditionally these communities have been mostly distinct, the lines between
them are increasingly blurred. In fact, fuzzing can be viewed as an instance of PBT where the property is fixed (e.g. the program does not crash) and PBT can be improved by extending its generation with coverage-guided capabilities. Such approaches have been tried recently with great success~\cite{FuzzChick, Zest}, and in the context of database testing, both SQLancer and \tool{} follow such a hybrid viewpoint.

\vspace{-2mm}


\subsection{SQLancer} \label{subsec:slancer}

SQLancer is a multi-year research project by Rigger et. al that set the bar for automated random testing of databases since its inception in 2020. It hosts an extensible core with adapters for many production DBMSs including but not limited to ClickHouse, Apache Datafusion, MySQL, PostgreSQL, and SQLite. Within the last five years, SQLancer has not only grown with respect to the breadth of databases it supports, but it has also widened its arsenal of oracles. It started with Pivoted Query Synthesis (PQS)~\cite{rigger.pqs}, a rather "simple" containment property over databases that has found at least 121 unique logic bugs in production databases. Two metamorphic oracles followed: Non-Optimizing Reference Engine Construction (NoREC)~\cite{rigger.norec}, which found 51 optimization bugs, and Ternary Logic Partitioning (TLP)~\cite{rigger.tlp}, which discovered 77 novel logic bugs.

SQLancer currently supports Query Plan Guidance (QPG)~\cite{rigger.qpg} for feedback-guided generation, Cardinality Estimation Restriction Testing (CERT)~\cite{rigger.cert} for finding performance bugs in DBMSs, Differential Query Plans (DQP)~\cite{rigger.dqp} for detecting bugs in join optimizations, and Constant Optimization Driven Database System Testing (CODDTest)~\cite{rigger.coddtest} for finding logic bugs in optimizations. As SQLancer focuses on testing large classes of behaviors across a variety of databases, each oracle amounts to a significant research contribution in a new research paper.

Integrating a new DBMS to SQLancer is a time-consuming
process.
%
At a minimum, SQLancer integration requires implementing AST connectors, generation APIs for the relevant queries, and the oracles to use for detecting bugs. SQLancer also has a notion of \emph{expected errors}, bugs that the users can deem as expected, so SQLancer does not report them as errors and continues testing.


\section{Specifying Correctness Oracles} \label{sec:correctness}

Rather than expecting database developers to modify SQLancer to suit their particular needs, we instead wanted to offer 
them flexible and extensible abstractions so that they test their database 
throughout its development.
\tool{} was created to provide such random testing infrastructure that evolves alongside the DBMS and produces actionable bug reports for the database developers. This evolution involves not only tailoring the input space of SQL generation to the currently implemented portion of \db{} for reducing false positives, but also tailoring the correctness oracles to new features as they are added, decreasing false negatives.

In this section, we introduce {\em generation actions}, 
an imperative formulation for properties that do not just describe \emph{what} the property tests, but \emph{how} to test it. We begin by using a very simple 
commutativity  property to convey the basic ideas and notation, then detail the oracles that were actually developed to test \db{}, and conclude the section with a description of our generation strategy.
For concreteness, let us consider the following equivalence relation leveraging the commutativity of $(\land)$: for any database $db$ and any two predicates $p$ and $q$ that can range over variables from the database, if we evaluate $SELECT$ $(p \land q)$  and $SELECT (q \land p)$ in $db$ they should yield equivalent results:
\begin{align*}
    \forall db, p, q. & variables(p) \subseteq db  \land variables(q) \subseteq db \\
    \implies & SELECT (p \land q) \equiv SELECT (q \land p)
\end{align*}
%
Testing such a property would entail generating an arbitrary database, two expressions $p$ and $q$ that only mention variables from that database, and then evaluating it repeatedly.
That is, we have precisely described \emph{what} we want to test, the exact conditions in which the test is valid, but left the \emph{how} to a database-agnostic framework. Instead of quantifying over $db, p, q$ and constraining them with post hoc relations, we propose a simple abstraction for describing \emph{how} $db, p, q$ are generated with respect to the constraints. We call this abstraction for describing properties \emph{Generation Actions (GA)}.

\begin{figure*}
    \centering
    \begin{subfigure}{0.45\textwidth}
    \begin{align*}
        &db: Database \\
        &t_1, t_2...t_n : Table \\
        &c_1, c_2...c_n : Column \\
        &p_1, p_2...p_n : Expression \\
        &r : Row \\
        &\forall db, t_1, t_2...t_n, c_1, c_2...c_n, p_1, p_2...p_n, r. \\
        &t_1, t_2...t_n \in db \\
        &\land c_1 \in t_1 \land c_2 \in t_2 ... \land c_n \in t_n \\
        &\land r.c_1 \in t_1.c_1 \land r.c_2 \in t_2.c_2 ... \land r.c_n \in t_n.c_n \\
        &\land p_1(r) = TRUE \land p_2(r) = TRUE ... \land p_n(r) = TRUE \\
        &\implies r \in SELECT * FROM \; t_1.c_1...t_n.c_n \; WHERE \; p_1 ... AND \; p_n
    \end{align*}
    \end{subfigure}
\hfill
\begin{subfigure}{0.45\textwidth}
\begin{gacode}
gen property db:
    (t1, t2) <- (pick db.tables, pick db.tables)
    (c1, c2) <- (pick t1.columns, pick t2.columns)
    (r1, r2) <- (gen row t1, gen row t2)
    r  := (r1.c1, r2.c2)
    
    ! INSERT INTO t1 VALUES r1
    ! INSERT INTO t2 VALUES r2
    
    p1 <- gen expression (t1, r1)
    p2 <- gen expression (t2, r2)
    
    ! RS := SELECT r FROM t1, t2 WHERE p1 AND p2
    ! assert(r in RS)
\end{gacode}
\end{subfigure}
\caption{Pivoted Query Synthesis as a universally quantified property (left) and as a generation action (right).}
\label{fig:pqs}
\end{figure*}





\setlength{\intextsep}{6pt}   
\setlength{\columnsep}{8pt}   
\begin{wrapfigure}{r}{0.22\textwidth}
\begin{gacode}
gen property db = 
  t <- pick db.tables
  c <- pick t.columns
  v <- genOf expression c.type
  p := t.c = v
  q <- gen expression (t,c)
  ! r1 := SELECT (p AND q)
  ! r2 := SELECT (q AND p)
  ! assert(r1 == r2)
\end{gacode}
\end{wrapfigure}

\noindent Each generation action is parameterized by the type it returns and the context it
can use (implemented as a trait in Rust). 
On the right, we express the same high-level commutativity property as a generation action that returns a \ga{property} parameterized by a context that contains a database \ga{db}.
Rather than independently generating $p$ and $q$, we fix some of the details of generation. We \ga{pick} an arbitrary table from the database, \ga{pick} a column from that table, and generate 
an expression \ga{v} of that column's type, before constructing an explicit equality check \ga{t.c = v}. For \ga{q}, we generate an arbitrary \ga{expression} that can refer to both \ga{t} and \ga{c}. We use $\leftarrow$ to bind the results
of generation and $:=$ for regular let-style binding.
We express interactions with the database, explicitly annotated with a $(!)$. In this example, the only interactions are defining the two different ways of conjuncting $p$ and $q$ and then \ga{assert}ing their equality, but can generally be a query or an assertion.


An interaction not shown in this example that we will use in our oracles is the ability to inject simulated {\em faults}. 
DBMSs interact heavily with the underlying OS, file system, and network, all of which have unpredictable and chaotic failure modes. As such, the same set of interactions (e.g. \sql{CREATE-INSERT-DELETE}) might result in different results depending on any invisible failures in the middle. FoundationDB~\cite{fdb} is heavily praised for its use of simulation testing, injecting simulated faults within an otherwise correctly working testing environment. We also introduced fault injection in \tool{} as part of the DSL for generation actions, as exemplified by the header initialization bug studied in \S \ref{case:headerinit}.

\subsection{Definitions of Oracles in \tool{}}

In this subsection, we go over definitions of five different oracles written as sequences of generation actions, three of which are reimplementations of existing oracles in SQLancer.
We start by defining the first of the three SQLancer oracles we implemented, Pivoted Query Synthesis (PQS)~\cite{rigger.pqs}, as a universally quantified proposition in Fig.~\ref{fig:pqs}. PQS states that given a set of tables in the database, \sql{SELECT}ing for a row constructed from the contents of those tables should contain the row. 

\begin{figure*}
\centering
\begin{subfigure}{0.45\textwidth}
\begin{gacode}
gen property db:
    t <- pick db.tables
    r <- pick t.rows
    p <- gen expression (t, r)
    ! RS1 = SELECT * FROM t WHERE p
    ! RS2 = SELECT p FROM t
    ! assert(RS1.length() == RS2.count(1))
\end{gacode}
\caption{Non-Optimizing Reference Engine\\Construction (NoREC)}
\label{fig:norec}
\end{subfigure}
\begin{subfigure}{0.45\textwidth}
\begin{gacode}
gen property db:
    t <- pick db.tables
    p <- gen expression (t)
    p' <- gen expression
    ! RS1 = SELECT * FROM t WHERE p
    ! RS2 = SELECT * FROM t WHERE p AND p' UNION ALL
            SELECT * FROM t WHERE p AND (NOT p') UNION ALL
            SELECT * FROM t WHERE p AND (p' is NULL)
    ! assert(RS1.length() == RS2.count(1))
\end{gacode}
\caption{Ternary Logic Partitioning (TLP) WHERE Extended}
\label{fig:tlp}
\end{subfigure}
\begin{subfigure}{0.45\textwidth}
\begin{gacode}
gen property db:
    t <- pick db.tables
    r <- pick t.rows
    p <- gen expression (t, r)
    ! DELETE * FROM t WHERE P
    ! RS = SELECT * FROM t WHERE p
    ! assert(r not in RS)
\end{gacode}
\caption{"Deletes rows should not be in the table"}
\label{fig:deleted_rows}
\end{subfigure}
\begin{subfigure}{0.45\textwidth}
\begin{gacode}
gen property db:
    s1 <- gen SELECT db.tables
    s2 <- gen SELECT db.tables
    ! RS3 = s1 UNION ALL s2
    ! assume(s1.first().length() == s1.first().length())
    ! assert(RS1.length() + RS2.length() == RS3.length())
\end{gacode}
\caption{"UNION ALL preserves cardinality"}
\label{fig:union_all}
\end{subfigure}
    
\caption{Definitions of Oracles as Generation Actions}
\label{fig:gen-actions}
\end{figure*}

Fig.~\ref{fig:pqs} shows two formulations of PQS side by side, the propositional formulation with universally quantified variables that define \emph{what} PQS is on the left and a GA formulation of PQS with two tables/columns as generation actions that define \emph{how} PQS is tested on the right, modeling the implementation of PQS that tests \db{} today.
The propositional formulation quantifies over a database and a sequence of tables, columns, and expressions, under the constraint that the expressions will return $TRUE$ when tested against a row $r$. 
The corresponding GA closely follows along, but ensures these constraints are satisfied by construction.
We start by picking tables \sql{t1, t2} from the state \sql{db.tables}, followed by picking columns from the respective table.
\sql{gen row t} and  \sql{gen expression} \sql{(t, r)} are dependent generation primitives, where the former generates a row based on the table \sql{t}, and the latter generates an expression that will evaluate to \sql{TRUE} for row \sql{r} of the table \sql{t}.

In Fig.~\ref{fig:gen-actions}, we provide definitions of SQLancer oracles as GAs in addition to the other oracles we implemented for \tool{}. 
Fig.~\ref{fig:norec} demonstrates how to write Non-Optimizing Reference Engine Construction Generation (NoREC)~\cite{rigger.norec} as a GA.
Fig.~\ref{fig:tlp} presents the GA for \texttt{WHERE Extended} case of Ternary Logic Partitioning (TLP)~\cite{rigger.tlp} oracle from SQLancer.
We can express other properties that are not present in SQLancer, including those fundamental to key-value stores such as \emph{Deleted rows should not be in the table} presented in Fig.~\ref{fig:deleted_rows}.
Properties, depending on how they are constructed, might sometimes be invalid. In Fig.~\ref{fig:union_all}, applying \sql{WHERE} or \sql{UNION ALL} operation for composing the results of multiple \sql{WHERE} queries require both sides of the \sql{WHERE}s to have the same number of columns.



\subsection{Query Generation}

We designed our query generation algorithm with three goals in mind.
First, we wanted each automatically generated database interaction to 
respect database state (e.g. we should not select from 
a table that does not exist unless a user-written generation action
explicitly calls for it). Second, we did not want to rely on querying 
the database itself to obtain the information necessary to ensure the first
goal---doing so would assume that those queries ran successfully, defeating the
purpose of testing. Finally, we wanted to allow database developers to specify,
in a lightweight manner, aspects  of the distribution of generated interactions that they deemed important, such as
read/write heavy queries.
To that end, we follow a generation-by-execution style approach~\cite{testifc},
in which a model of the database as a key-value store is updated during generation as a shadow state. Unlike traditional model-based properties~\cite{htsi},
this model is not used for differential testing, only for keeping track of 
the relevant information for correct generation.
Algorithm~\ref{alg:gen} sketches our approach.

\begin{algorithm}
\caption{\tool{} Generation Algorithm}\label{alg:gen}
\begin{algorithmic}
\State $e \gets \{read: R, write: W, create: C\}$ \Comment{Workload distribution}
\State $interactions \gets []$
\State $st \gets \{read: 0, write: 0, create: 0, tables: []\}$
\While{$N > 0$}
    \State $i \gets gen \; Interaction (st, e)$
    \State $i.shadow(state)$ \Comment{Update the shadow state}
    \State $interactions.push(i)$
    \State $N \gets N - 1$
\EndWhile
\end{algorithmic}
\end{algorithm}

Database developers can specify three parameters, $R, W,$ and $C$, describing the proportion of read, write, and create instructions that should appear in the generated interactions. The algorithm iteratively generates one or more interactions compatible with the current shadow state, updates the state to account for the new interactions, and repeats until enough interactions have been generated.

%
%


SQLancer does not keep a separate state~\cite{rigger.pqs}, but instead uses the database APIs for querying the current state, such as the table names in \texttt{sqlite\_master}, due to the implementation effort for the shadow model. We opted for the shadow state because it allows for complex reasoning over the database state for constructing arbitrary queries and properties, and because it gives us a canonical property over the database state: the shadow state is identical to the database at any given moment.

\section{Evaluation}\label{sec:eval}

In order to evaluate the effectiveness of \tool{}, we explored the answers to the following questions:

\begin{itemize}
    \item RQ1: Does \tool{} find bugs in \db{}?
    \item RQ2: How does \tool{} compare to the state-of-the-art?
\end{itemize}

\subsection{RQ1: Does \tool{} find bugs in \db{}?} \label{subsec:bugs}

\setlength{\dbltextfloatsep}{6pt}
\begin{small}
\begin{table*}[h!]
    \centering
    \begin{tabular}{ | l | l | l | l |}
\hline
\textbf{Bug Id} & \textbf{Description} & \textbf{Oracle} & \textbf{Module} \\
\hline

466 & TRUE not accepted as catch-all predicate & No Error & Query Compiler  \\
\hline  
548 & Infinite loop when checkpointing on Linux & No Infinite Loop & I/O Subsystem  \\
\hline  
629 & Query Optimizer broke TRUE in predicates & No Panic & Query Optimizer  \\
\hline  
662 & SELECT with nested Boolean expressions sometimes gave no results & PQS & Query Optimizer  \\
\hline  
681 & Storage Engine (B-tree) insert caused subtract with overflow & No Panic & Storage Engine (B-tree)  \\
\hline  
682 & Faulty recursive binop logic caused SELECT to miss rows & PQS & Query Compiler  \\
\hline  
924 & Storage Engine (B-tree) balancing caused page corruption when deleting & No Panic & Storage Engine (B-tree)  \\
\hline  
1040 & LIKE operator did not work for non-text values & No Panic & Query Executor  \\
\hline  
1203 & Storage Engine (B-tree) balancing error & No Panic & Storage Engine (B-tree)  \\
\hline  
1629 & Storage Engine (B-tree) cell updates caused infinite loop in UPDATE & No Infinite Loop & Storage Engine (B-tree)  \\
\hline  
1734 & DELETE did not emit conditional jumps if WHERE term was constant & Delete-Select & Query Compiler  \\
\hline  
1815 & Storage Engine (B-tree) failed to balance when insert caused cell overflow & No Panic & Storage Engine (B-tree)  \\
\hline  
1818 & Always read DB header and schema from file instead of memory page & No Panic & Page Manager  \\
\hline  
1975 & Storage Engine (B-tree) expected table or index leaf page & No Panic & Storage Engine (B-tree)  \\
\hline  
1991 & Use after free when validating B-tree balance & No Panic & Storage Engine (B-tree)  \\
\hline  
2024 & SELECT ... LIMIT resulted in different rows from SQLite  & Differential & Query Executor  \\
\hline  
2026 & UPDATE then SELECT resulted in different rows from SQLite  & Differential & Query Executor  \\
\hline  
2047 & Overflow cell with divider cell was not found due to faulty validation& No Panic & Storage Engine (B-tree)  \\
\hline 
2074 & SELECT hung with long text, CacheFull error & No Panic & Page Manager  \\
\hline  
2075 & Large table with 128 columns handled incorrectly & No Panic & Page Manager  \\
\hline  
2088 & Incorrect record header size calculation & No Panic & Page Manager  \\
\hline  
2106 & Interior node replacement caused self-reference when depth exceeded 2 & No Panic & Storage Engine (B-tree)  \\
\hline  
2116 & Advance after post-delete balancing did not advance B-tree cursor & No Panic & Storage Engine (B-tree)  \\
\hline 
\end{tabular}
    \caption{List of confirmed unique \db{} bugs found and reported by \tool{}}
    \label{tab:bugs}
\end{table*}
\end{small}

We have collected over 23 confirmed and fixed unique bugs found by \tool{} in \db{} over the course its development, evolving in complexity as \db{} grows. Table~\ref{tab:bugs} is the list of confirmed bugs, along with the changes to the testing infrastructure that resulted in the discovery.
The bugs range from simple parser level bugs that could be classified as minor problems, to severe logic bugs deep in the core functionality of the database that could result in data loss. 
We choose three representative bugs as case studies, detailing the conditions in which they were discovered and triggered. 
The cases demonstrate the diversity of bugs found in \db{} with respect to their origin (the bytecode compiler, the B-tree index, and the database header), to how they were discovered and reported (automated report, ourselves, and the developers), and to the oracles used (simple deletion properties, large indexed table generation, and developer-written
properties).

\paragraph{Case 1: DELETE not emitting constant Where terms.} 

The first bug lies in the compiler of \db{}. \db{} translates SQL statements into bytecode and then runs that bytecode in a virtual machine (this is the same approach that SQLite uses~\cite{sqlite}).
The bytecode compilation is error-prone, as small changes to the compiled SQL expression can greatly affect the generated bytecode. 


The bug stemmed from an error in the bytecode compilation of \sql{DELETE}, specifically in the case of constant expressions in the \sql{WHERE} terms. Constant expressions can be compiled to run before the main execution loop of the query, but instead the compilation step just skipped them. This caused \sql{DELETE} with a constant that evaluates to \sql{FALSE} to be incorrectly executed.



This bug was automatically reported by \tool{} running in \db{} CI and fixed within the week. The bug was discovered as a result of our the PQS implementation in \db{} augmented with validity preserving queries between the \sql{INSERT} and \sql{SELECT} statements. The random tester added a naive \sql{DELETE} operation that should not have affected the results of the \sql{SELECT}, keeping the containment property intact, but the assertion that the inserted row should be in the result of the \sql{SELECT} failed, triggering the bug report.

\paragraph{Case 2: Interior node replacement caused self-reference when depth exceeded 2.} The second bug manifested in the balancing logic of the B-tree implementation of \db{}, but only 
because the generated interactions were paced to match the state of the \db{} development.
When \db{} added database indexes as an experimental feature, we extended generation to support statements such as \sql{SELECT DISTINCT}, \sql{CREATE INDEX}, and compound operators such as \sql{UNION} or \sql{UNION ALL} to the space of generated inputs, and started generating larger tables to take advantage of the indexes. This resulted in triggering a crash failure in the core B-tree data structure that could result in data loss or corruption if not fixed. We reported this bug to the \db{} developers, and the bug is currently fixed.

\paragraph{Case 3: Database Header Initialization.}\label{case:headerinit} The third selected bug highlights the importance of considering fault scenarios. SQLite uses write-ahead logs (WAL) in order to provide non-blocking reads and ACID transactions in the database. This strategy incurs an additional complexity to the persistent state of the DBMS as the database file might be out-of-date for certain operations, requiring the DBMS to read such information from WAL.
This bug is the result of such a situation, where metadata like database size and schema is out-of-date at the time of reading, leading to incorrect operations. \tool{} found this bug through an assertion failure in the freelist structure in the database header.
As a result, the developer reporting the original bug has written the first example of a \emph{regression property}, added a new fault primitive \sql{ReopenDatabase} that closes down existing connections with the database and reopens them later, enabling the detection of any future bugs that might be a result of an error in the persistent state logic of \db{}.


\subsection{RQ2: How does \tool{} compare to SQLancer?}

\begin{figure*}[h!]
    \centering
    \includegraphics[width=0.85\linewidth]{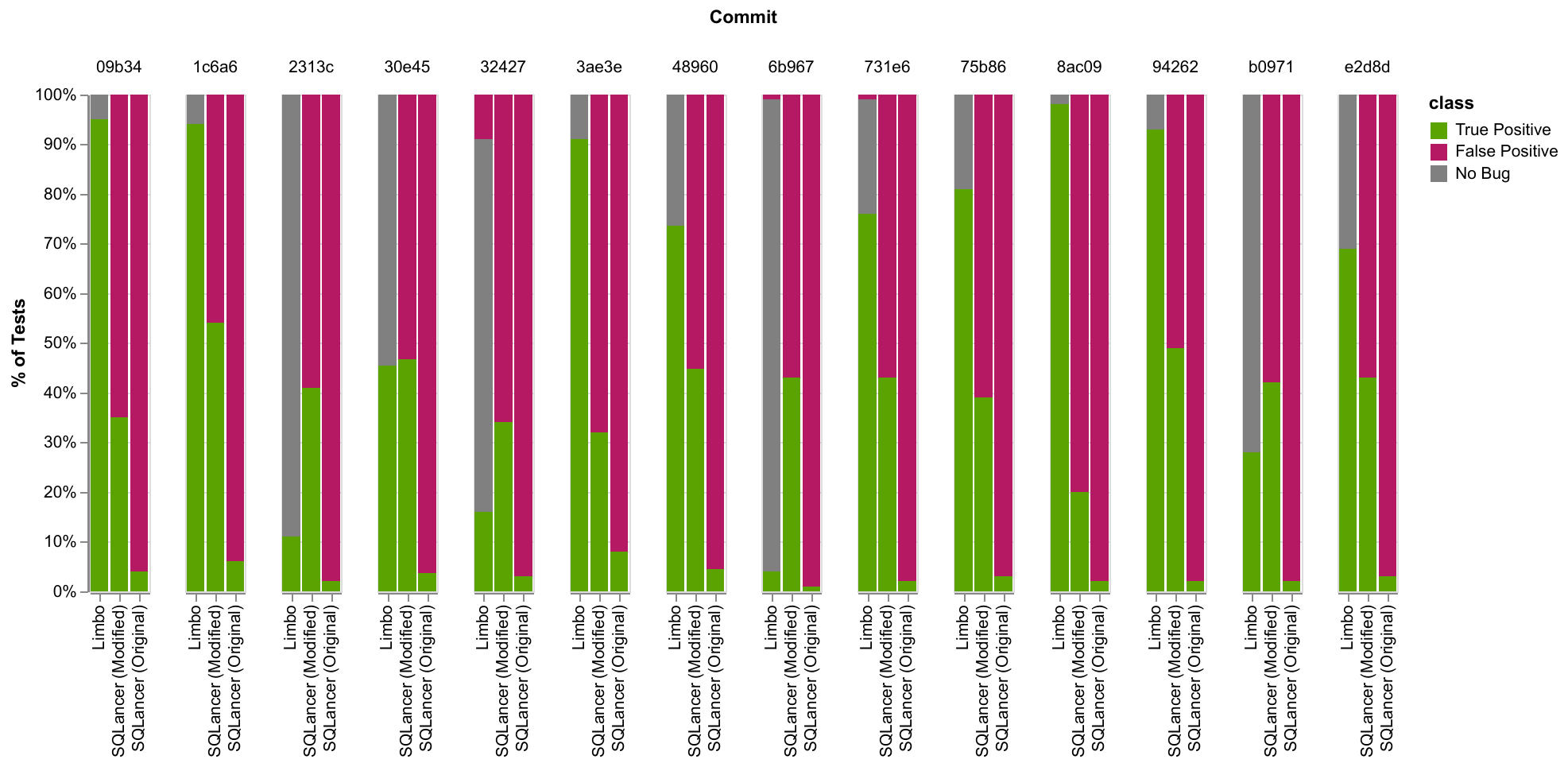}
    \caption{Comparison of True Positives, False Positives and No Bugs for \tool{}, SQLancer-SQLite, SQLancer-\db{}
    }
    \label{fig:comparison}
\end{figure*}

We answer RQ2 by comparing the rate of true positives and false positives reports from three different random testing configurations: \tool{}, the default SQLite3 integration of SQLancer with pragmas disabled (denoted SQLancer-SQLite), and the fork of a work-in-progress \db{} integration of SQLancer by a \db{} contributor (denoted SQLancer-\db{}). Pragmas had to be removed for SQLancer-SQLite because most of them were unimplemented in Turso, causing all reported bugs to be false positives in our testing.
We define an actionable bug report from the tools as a sequence of SQL statements that \emph{would have been} submitted and confirmed as a bug at the time of its original report. We used the 23 bug reports we analyzed as the ground truth, finding commits which are known to have bugs. The bugs span over 20 \db{} commits, 6 of which do not provide Java bindings necessary for connecting with SQLancer. We ran all three tools on the remaining 14 commits in the commit history.
We analyzed the results of each run and separated them into three bins: \emph{True Positives} that would have been submitted and confirmed as bugs, \emph{False Positives} that are reported as failures by the tools but do not constitute bugs as they are unimplemented features, and \emph{No Bugs} that did not report a failure, provided in the stacked bar chart in Fig.~\ref{fig:comparison}.
SQLancer-SQLite achieves a true positive rate below 10\% across all 14 commits. SQLancer-\db{} performs better, with a 42\% true positive rate, but still yields many more false positives than \tool{} (58\% vs.\ under 1\%). \tool{} also finds substantially more actionable bugs: 25 unique bugs, compared with six found by the two SQLancer configurations combined (\S\ref{app:sqlancer-bugs}).
This does not mean that \tool{} has better generators or oracles than SQLancer. Rather, it shows that for rapidly evolving databases with many missing features, database-integrated testing yields more actionable bugs.

\section{Discussion}

\tool{} is a paradigm for testing databases during development, even though we use \tool{} the paradigm and \tool{} for \db{} interchangeably throughout the paper. We compare the performance of \tool{} for \db{} against SQLancer with minimal modifications, which begets some questions worth discussing.

\emph{Why not integrate SQLancer to the development process instead of developing a random testing framework from scratch?} As we briefly discussed earlier, SQLancer is not designed for the constant evolution of the random testing infrastructure along with the project. It is a project with its own trajectory, development, new algorithms, generators, and oracles.
\tool{} aims to empower the database developers themselves by giving them a mechanism to test the database, instead of giving them bugs to solve. There are additional practical barriers; Java bindings for \db{} are not complete at the moment, so we had to fix bugs in the SQLancer-\db{} integration implemented by one of the \db{} developers that relied on incomplete features that silently failed, panics in \db{} caused SQLancer to terminate, so it was not possible to minimize the counterexamples with SQLancer reducers without further changes.

\emph{Given that SQLancer supports expected errors for reducing false positives, could we use expected errors for known bugs and false positives we found to discover more bugs?} We could use expected errors, and we would have discovered more bugs in the process. In terms of any static evaluation target, tweaking SQLancer by progressively finding more bugs as found ones are marked as expected, it is always possible to make SQLancer find more bugs. The point, however, is that constant modification is not the expected and supported mode of operation when using SQLancer. Although it is phenomenal for off-the-shelf usage, its capabilities unfortunately act as a disadvantage against its use in actively developed projects with many missing features. We have shown that a project with a much smaller scope can adopt the ideas of input generation and oracles in SQLancer as well as other related work on database testing such as Apollo~\cite{apollo} or Thanos~\cite{thanos} and demonstrated that database developers can turn their domain expertise into writing properties as presented in the third case study in \S \ref{subsec:bugs}.
We also observed that many crash failures are the result of inline assertions in \db{}. These crashes imply logic bugs as they contradict invariants specified by the developers, blurring the distinction between crash failures and logic bugs in the literature. Such assertions do not replace SQLancer oracles or \db{} properties because they do not have the ability to follow values through execution in a holistic way as properties do, so they can only reason about local invariants.

\emph{Role in \db{}.}
As a last point of discussion, we would like to clarify our role in the implementation of \tool{} in \db{}, in order to fully credit the open source contributors for their work. At the time we started working on \db{}, it had a small, unstructured random testing infrastructure that overwhelmingly focused on being able to support Deterministic Simulation Testing~\cite{fdb}. I/O was implemented in a way that could be easily simulated, which allowed for \sql{FAULT}s to be integrated in the generated interactions. 
We have taken a role as an external open source contributor to the project, gradually proposing improvements to random testing infrastructure such as the Generation Actions DSL, most of the existing properties, stateful random generation, shrinking, additional oracles such as differential testing against SQLite and determinism checking, and have implemented such proposals. The infrastructure has grown outside of our control at times, producing regressions through the development, along with many improvements by the maintainers and contributors to the project that contributed to the list of bugs found in Table~\ref{tab:bugs}. The SQLancer-\db{} integration was almost entirely developed by one contributor, which we have used in our evaluations with small changes to their code.

\emph{Related work on adaptive DBMS testing.} Although we compared \tool{} mainly against SQLancer as the basis of our oracles, other systems such as SQLRight~\cite{sqlright}, Griffin~\cite{griffin}, and SQLancer++~\cite{zhong2026scalingautomateddatabasetesting} also adapt testing to DBMS behavior. However, these systems primarily approach adaptation from the perspective of an external testing framework, with evaluations centered on broad effectiveness across multiple databases and minimal manual intervention. In contrast, \tool{} studies adaptation as a database-integrated process: the testing infrastructure evolves together with a single DBMS under development, and developers directly shape generators, properties, and fault models as the implementation changes.

\section{Conclusion and Future Work}\label{sec:futurework}

In this paper, we introduced \tool{}, a new paradigm and concrete implementation for testing databases during development. By tightly integrating the testing infrastructure with the DBMS itself, \tool{} avoids many of the limitations that off-the-shelf testing frameworks like SQLancer face when applied to rapidly evolving systems with incomplete feature sets. Our framework provides developers with tools to express custom correctness properties via generation actions and ensures state-aware query generation.
We implemented \tool{} for \db{} and found 23 confirmed bugs with minimal false positives. In the same setting, SQLancer-based baselines produced substantially more false positives and fewer actionable reports.
Our experience shows that when databases are still under heavy development, flexible, integrated testing approaches that evolve alongside the system are significantly more productive and developer-friendly than external, comprehensive tools. Moreover, our approach fosters collaboration between testing infrastructure and DBMS development by giving developers intuitive ways to encode domain-specific knowledge as properties.


\tool{} bridges the gap between random testing and practical debugging by empowering developers to guide, understand, and act on test results. We believe this paradigm has broad applicability beyond \db{} and offers a promising path forward for testing other evolving and complex systems.

\appendix

\section{Appendix} \label{sec:appendix}

\subsection{Bugs Found by SQLancer} \label{app:sqlancer-bugs}

\begin{itemize}
    \item (\sql{UPDATE t SET (c0, c0)=(0, 0)}): This expression results in a "Column specified more than once" error that has not been fixed at the time of our submission.
    \item (\sql{SELECT (0x0)}): There was a parsing error for hexadecimals that resulted in an "invalid float literal" error that has since been fixed before our experiments with SQLancer.
    \item (\sql{SELECT * FROM t WHERE c0 GLOB c0}): There was a logic bug in the \sql{GLOB} that panicked when the values passed were not \sql{TEXT}. Executing this statement after inserting a \sql{NULL} triggered the bug, which was fixed soon after we reported it.  
    \item (\sql{INSERT INTO t VALUES ("a")}): Using double quotes for string literals is discouraged, and SQLite3 even provides a runtime flag for disabling it. We reported this bug as \db{} panicked for the provided statement.
    \item (\sql{INSERT INTO t VALUES ((0 BETWEEN 0 AND 0)), (0)}): Constant \sql{BETWEEN} expressions were not  rewritten when used within \sql{INSERT}, but the query compiler expected them to be rewritten. We reported this bug as \db{} panicked for the provided statement.
    \item (\sql{INSERT INTO t(c2, c0) VALUES (0, 0), (0, 0)}): There was a bug in calculating the column indexes when inserting values to a table in reverse order. SQLancer discovered the bug when the initial table \sql{t} had a \sql{NOT NULL} clause for the column \sql{c0}, which promptly failed after executing the statement. We reported this bug after manually inspecting the result of the failure.
\end{itemize}

In all six cases, we see the benefit of generating the entire possible input space for SQL dialect of SQLite, which \tool{} does not, hence missing such bugs. Our evaluation demonstrates the benefit of such comprehensive generation, but also shows that without tailoring to the database, the produced bug reports will be overwhelmingly dominated by false positives and reproductions of existing bugs yet to be fixed.

\bibliographystyle{acm}
\bibliography{references}
\end{document}